\documentclass{emulateapj}
\usepackage{txfonts}
\usepackage{graphicx}
\usepackage{subfigure}
\usepackage{multirow}
\usepackage{url}
   
\def\lsim{\mathrel{\rlap{\lower4pt\hbox{\hskip1pt$\sim$}}
    \raise1pt\hbox{$<$}}}                
\def\gsim{\mathrel{\rlap{\lower4pt\hbox{\hskip1pt$\sim$}}
    \raise1pt\hbox{$>$}}}                

\shorttitle{Unusually Wide M-dwarf Binaries}
\shortauthors{N.M. Law et al.}

\begin{document}

\title{The High-Order-Multiplicity of Unusually Wide M-dwarf Binaries: Eleven New Triple and Quadruple Systems}

\author{N.M. Law\altaffilmark{1}, S. Dhital\altaffilmark{2}, A. Kraus\altaffilmark{3}, K. G. Stassun\altaffilmark{2,4}, A. A. West\altaffilmark{5}}
\altaffiltext{1}{Dunlap Fellow, Dunlap Institute for Astronomy and Astrophysics, University of
Toronto, 50 St. George Street, Toronto M5S 3H4, Ontario, Canada}
\altaffiltext{2}{Department of Physics \& Astronomy, Vanderbilt University, 6301 Stevenson Center, Nashville, TN, 37235, USA}
\altaffiltext{3}{Hubble Fellow, Institute for Astronomy, University of Hawaii, 2680 Woodlawn Drive, Honolulu, HI, 96822, USA}
\altaffiltext{4}{Department of Physics, Fisk University, 1000 17th Ave. N., Nashville, TN 37208}
\altaffiltext{5}{Department of Astronomy, Boston University, 725 Commonwealth Avenue, Boston, MA 02215, USA}

\begin{abstract}
M-dwarfs in extremely wide binary systems are very rare, and may thus have different formation processes from those found as single stars or close binaries in the field. In this paper we search for close companions to a new sample of 36 extremely wide M-dwarf binaries, covering a spectral type range of M1 to M5 and a separation range of 600 - 6500 AU. We discover 10 new triple systems and one new quadruple system. We carefully account for selection effects including proper motion, magnitude limits, the detection of close binaries in the SDSS, and other sample biases. The bias-corrected total high-order-multiple fraction is ${45}^{+18}_{-16}$\% and the bias-corrected incidence of quadruple systems is $<$5\%, both statistically compatible with that found for the more common close M-dwarf multiple systems. Almost all the detected companions have similar masses to their primaries, although two very low mass companions, including a candidate brown dwarf, are found at relatively large separations. We find that the close-binary separation distribution is strongly peaked towards $<$30AU separations. There is marginally significant evidence for a change in high-order M-dwarf multiplicity with binding energy and total mass. We also find 2$\sigma$ evidence of an unexpected increased high-order-multiple fraction for the widest targets in our survey, with a high-order-multiple fraction of ${21}^{+17}_{-7}$\% for systems with separations up to 2000AU, compared to ${77}^{+9}_{-22}$\% for systems with separations $>$4000AU. These results suggest that the very widest M-dwarf binary systems need higher masses to form or to survive.
\end{abstract}


\maketitle

\section{Introduction}

Multiple star systems offer a powerful way to constrain the processes of star formation. The distributions of companion masses, orbital radii and thus binding energies provide important clues to the systems' formation processes. A commonly invoked model for very low mass star formation is fragmentation of the initial molecular cloud core, followed by ejection of the low mass stellar embryos before mass accretion has completed -- the ejection hypothesis (\citealt{Reipurth2001}, \citealt{Close2003}, \citealt{Siegler2005}, \citealt{Konopacky2007}).  The model successfully predicts the observed field close binary frequency and median separation (e.g. \citealt{Bate2009}), and predicts that essentially no binaries with low binding energies are expected to survive the ejection. 

Nevertheless, several extremely wide M-dwarf binaries have been detected in the field (\citealt{PhanBao2006}, \citealt{Artigau2007}, \citealt{Caballero2007}, \citealt{Artigau2007}, \citealt{Law2008}, \citealt{Radigan2009}, \citealt{Zhang2010}, \citealt{Faherty2010}). With total system masses of around \mbox{0.2 $\rm{M_{\odot}}$} and orbital radii of thousands of AU, these systems have a very low binding energy compared to the limits set by the ejection hypothesis. 

On the other hand, there is a sharp binding energy cutoff observed for almost all field VLM binaries (e.g. \citealt{Close2003}, \citealt{Konopacky2007}), so perhaps the ejection hypothesis is not invalidated by the few known very low binding energy systems. It is plausible that these "uncommon" systems represent a different mode of wide binary production, such as mutual ejection \citep{Kouwenhoven2010}. If these unusually wide systems are indeed formed in a different way, the higher-order multiplicity statistics of the systems may be different from that of systems with larger binding energies.

In addition to binding energy considerations for VLM stars, empirical limits on binary separation have also been found among higher-mass M-dwarfs: $\rm \log a_{max} = 3.3 M_{tot} + 1.1$ for total masses greater than 0.3${\rm M_{\odot}}$ \citep{Reid2001}, and for lower masses $\rm a_{max} = 1400 M_{tot}^{2}$ \citep{Burgasser2003}, where $\rm{M_{tot}}$ is the total system mass in solar masses and $\rm{a_{max}}$ is the maximum separation in AU. Systems that violate these limits could be considered ``unusual'', and may similarly have different multiplicity properties from the much more common close M-dwarf binaries. 

\begin{figure}
  \centering
  \resizebox{1.0\columnwidth}{!}
   {
	\includegraphics{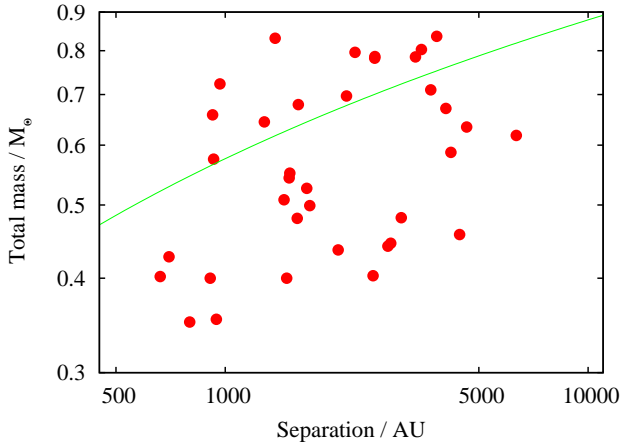}
   }
   \caption{The separations and masses of our wide binary sample. Masses are interpolated from the spectral type / mass relations in \citealt{Kraus2007a}. The empirical field binary separation cutoff from \citet{Reid2001} is shown as a solid line; we define a system as ``unusually wide'' if it falls below this line.}
   \label{FIG:mass_sep}
\end{figure}

\begin{figure*}[]
  \centering
	\subfigure{\resizebox{0.45\textwidth}{!}{{\includegraphics{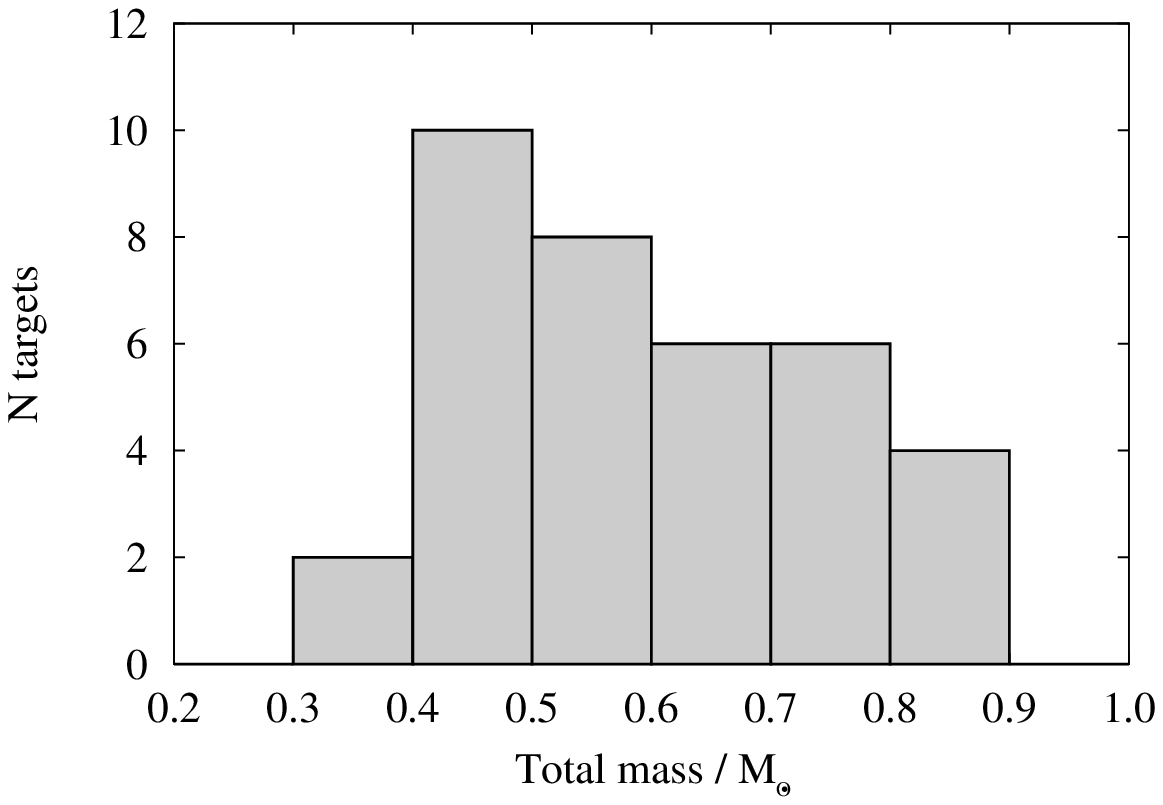}}}}\hspace{0.15in}
	\subfigure{\resizebox{0.45\textwidth}{!}{{\includegraphics{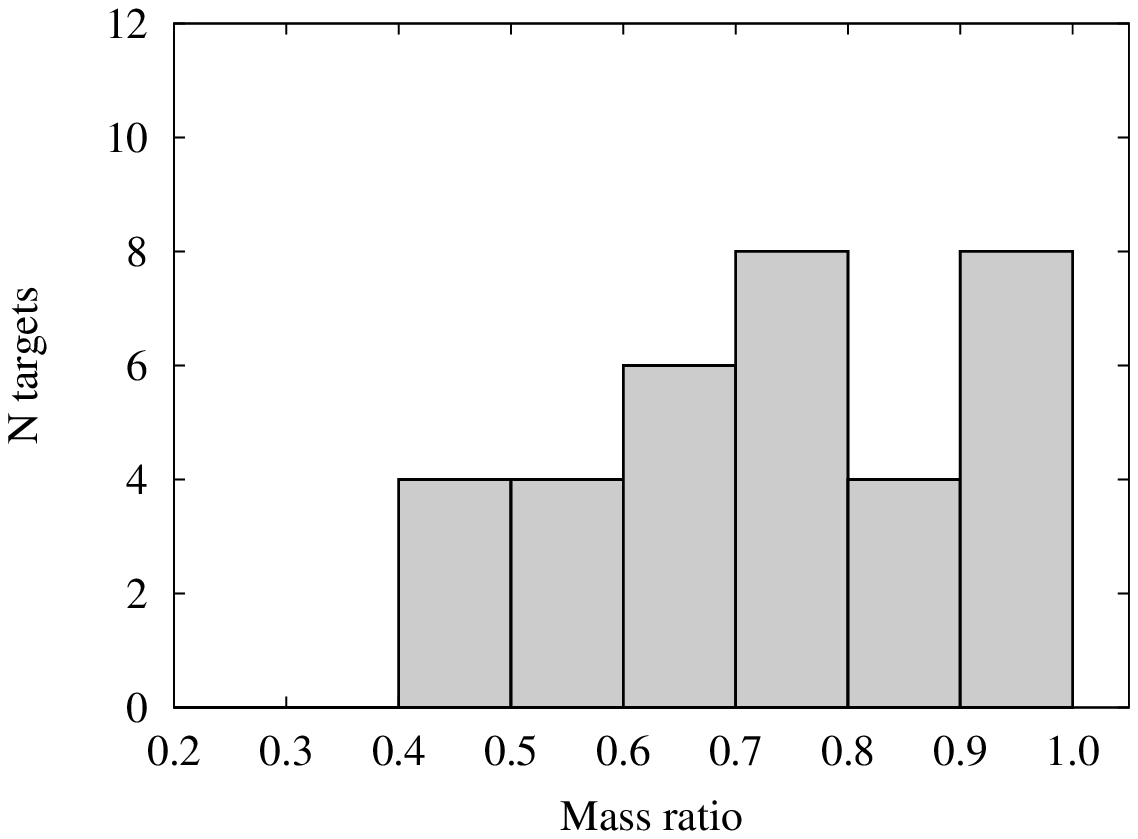}}}}
	\subfigure{\resizebox{0.45\textwidth}{!}{{\includegraphics{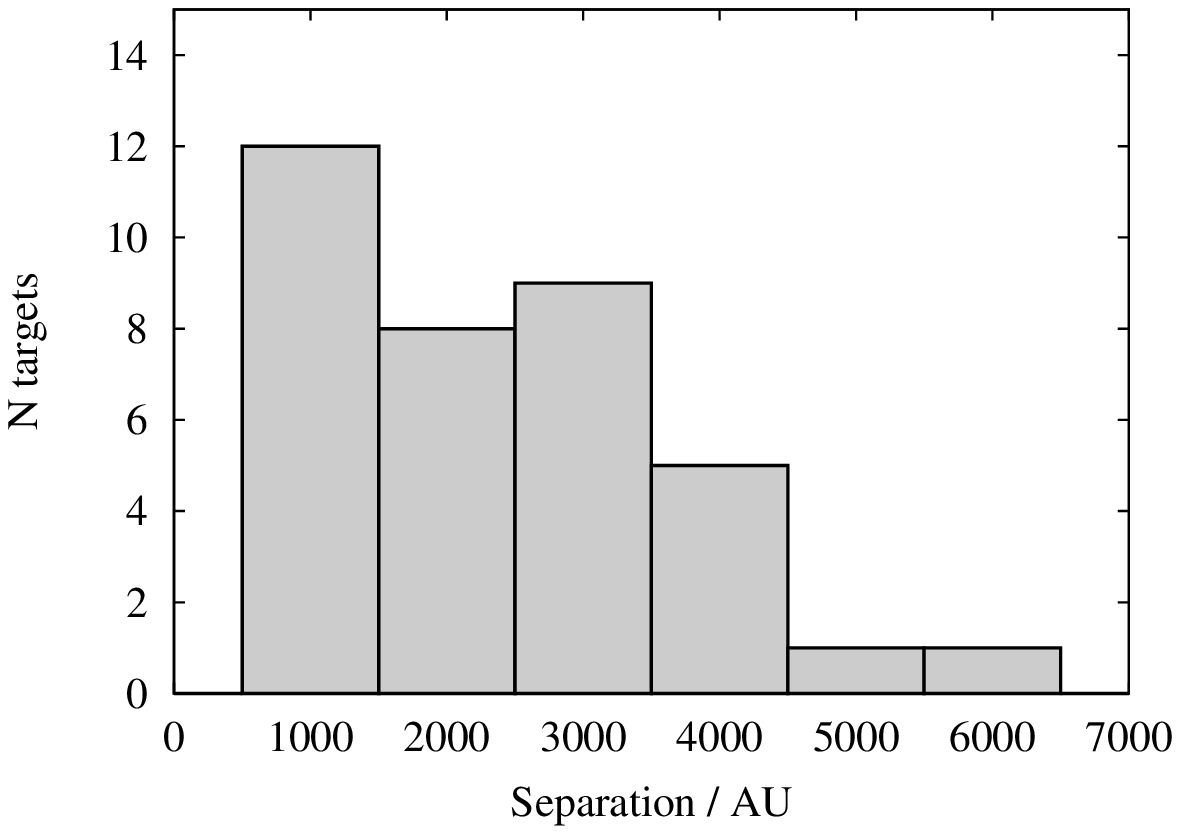}}}}\hspace{0.15in}
	\subfigure{\resizebox{0.45\textwidth}{!}{{\includegraphics{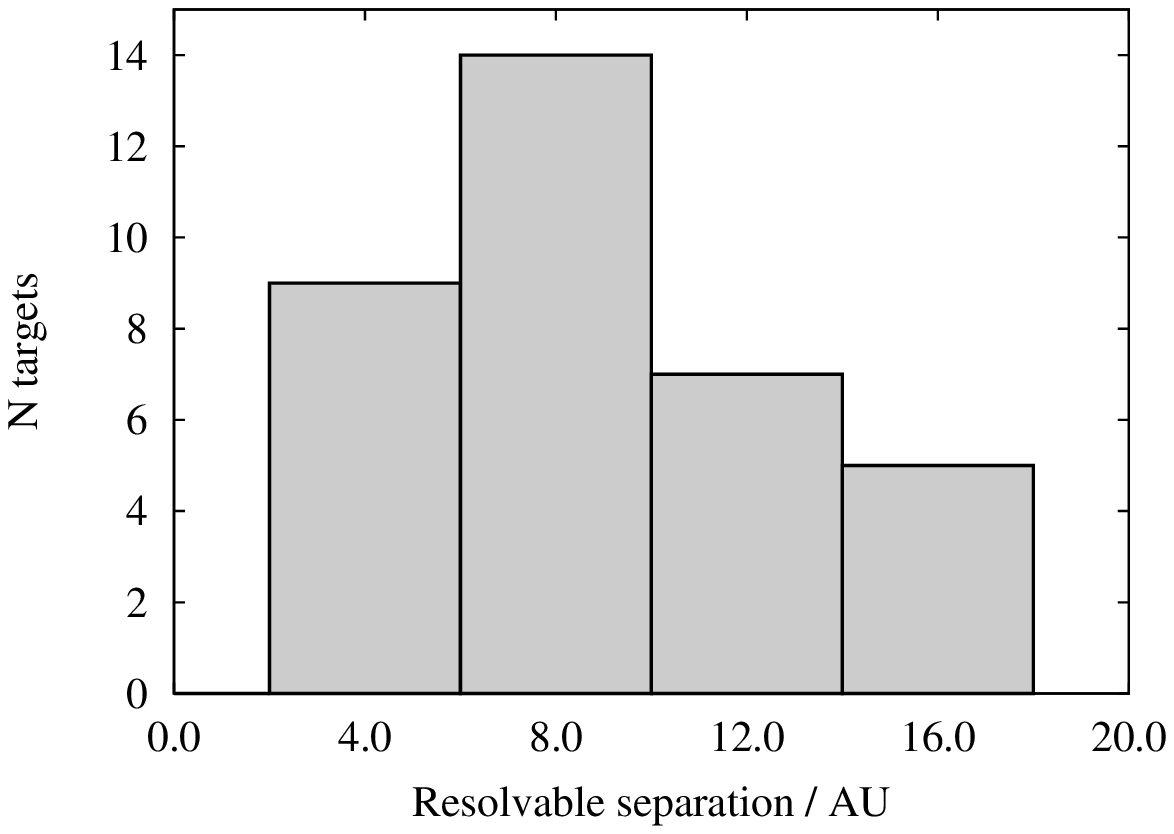}}}}
   \caption{The properties of the target sample based on SDSS photometry (and thus not including the effects of close companions). Clockwise from top left: 1) the total mass of the binaries; 2) the mass ratios of the binaries; 3) the minimum resolvable separation (Rayleigh criterion for each telescope); and 4) the projected separation of the binary components.}
   \label{FIG:target_hists}
\end{figure*}

The total masses of such unusually wide systems are generally inferred from seeing-limited photometric or spectroscopic fits to determine the spectral types of the bodies making up the binary. The discovery of closer companions can thus dramatically alter the inferred total masses of the systems, and therefore the expected applicable separation limits. If a significant fraction of apparently wide binaries are actually much higher mass high-order multiple systems, the inferences on their formation environments must be correspondingly altered. In addition, wide binary systems which survive the formation process may have experienced smaller dynamic influences on their evolution. It is conceivable that lower-mass substellar or even planetary companions may be more common in such systems.

New proper motion surveys based on the Sloan Digital Sky Survey \citep[SDSS; ][]{York2000}  are finding large samples of M-dwarf binaries which apparently violate the \citet{Reid2001} \& \citet{Burgasser2003} empirical limits (e.g. SLoWPoKES; \citealt{Dhital2010}). For the first time, these new samples allow us to undertake detailed statistical studies of the systems' multiplicity properties.

In this paper we target 36 extremely wide binaries, chosen to cover the mass and separation range outlined by the \citet{Reid2001} limits. Using Keck Laser Guide Star Adaptive Optics (LGS-AO; \citealt{Wizinowich2006}) and Palomar natural guide star adaptive optics \citep{Troy2000}, we searched each target system for companions at separations ranging from 4 to 100 AU, and at contrast ratios sufficient to detect brown dwarf companions.

We detail the construction of the new sample of extremely wide M-dwarf binaries in Section \ref{sample}. The observations and data reductions are described in Section \ref{obs}. Sections \ref{results} and \ref{discussion} detail and discuss the survey results, and we conclude in Section \ref{conclusions}.

\section{Construction of the Wide M-Dwarf Binary Sample}
\label{sample}
The targets were selected from a preliminary version of the SLoWPoKES catalog, a sample of wide ($>$ 500 AU), low-mass (mid-K--mid-M) common proper motion (CPM) binary systems in the SDSS Data Release 7 \citep[DR7;][]{Abazajian2009}. We briefly describe the selection algorithm here; the full selection process is detailed in \citet{Dhital2010}. 

The SDSS DR7 photometric catalog has more than 180 million stellar sources, of which $\sim$109 million are low-mass mid-K--late-M dwarfs; many sources also have measured proper motions in the SDSS/USNO-B matched catalog \citep{Munn2004, Munn2008}. To identify CPM pairs from this large database \citet{Dhital2010} used a relatively high proper motion sample ($\mu\geq$ 40 mas~yr$^{-1}$) to avoid high field contamination. Candidate binaries were identified at angular separations of 7--180$\arcsec$, very conservatively requiring photometric distances and component proper motions to individually match within 1-$\sigma$ [0.3--0.4 magnitudes \citep{Bochanski2010} in photometric distance modulus, and $\sim$2.5--5 mas~yr$^{-1}$ in proper motion \citep{Munn2004, Munn2008}]. A detailed galactic model was used to quantify the probability of a random alignment, and only pairs with a $<5\%$ chance alignment probability were accepted. The final SLoWPoKES catalog contains 1336 very wide, low-mass CPM pairs.

To generate our close-companion-search target list from the SLoWPoKES catalog we required 1) both components to be M-dwarfs and later than M1; 2) the components to have an angular separation $<$40\arcsec (to allow simultaneous imaging of both components); and 3) at least one component to be brighter than r$\approx$17. The latter requirement was imposed due to poor conditions during all our LGS AO observations, and effectively limited the primaries of our targets to be earlier than M5. The resulting 36-binary sample (Table \ref{tab:targets}) covers a total range of 0.35 - 0.85 $\rm M_{\odot}$ in mass and 600 - 6500 AU in separation (Figure \ref{FIG:target_hists}). 

\begin{figure}
  \centering
  \resizebox{1.0\columnwidth}{!}
   {
	\includegraphics{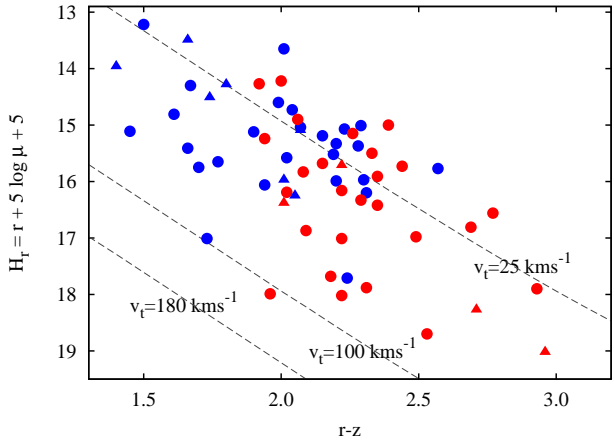}
   }
   \caption{A reduced proper motion plot for our targets. Primaries are blue and secondaries are red; the targets which are in fact high-order-multiples are plotted as triangles. Lines of tangential velocity for the disk, halo and an intermediate value are plotted for comparison.}
   \label{fig:rpm}
\end{figure}

Using the \citet{Dhital2010} galactic model and the galactic positions of our targets, we estimate that our target binaries each have a 96\% chance of being members of the thin disk, a $\approx$4\% chance of being part of the thick disk, and a $<$1\% chance of being halo members. A reduced proper motion diagram supports this conclusion (Figure \ref{fig:rpm}), although the spread in estimated tangential velocities suggests that a few of the targets may be members of the thick disk.

Our sample is bisected by the \citet{Reid2001} field M-dwarf maximum separation distribution (Figure \ref{FIG:mass_sep}). 23 of the 36 targets are considered unusually wide by the \citet{Reid2001} definition. We note that there is no clear reduction in our target numbers beyond the \citet{Reid2001} cutoff, suggesting that the new binary surveys now available may require the revisiting of this criterion (e.g. \citet{Dhital2010}). None of our targets violate the suggested alternate Jeans-length based criteria on the system binding energy \citep{Zuckerman2009, Faherty2010}, although all are close to the minimum allowed binding energies and allow us to search for a variation in multiplicity properties across a wide range of masses and separations.

\begin{figure}
  \centering
  \resizebox{1.0\columnwidth}{!}
   {
	\includegraphics{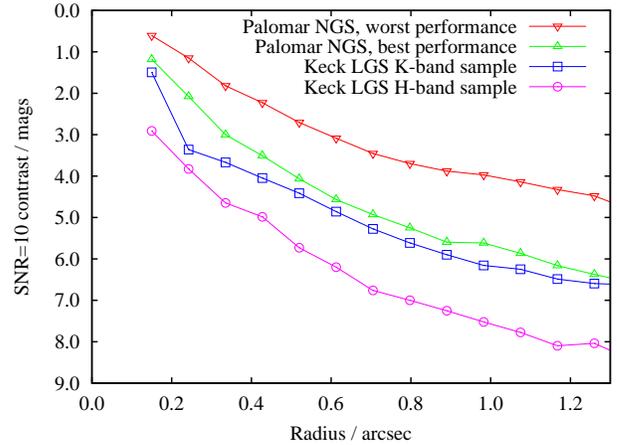}
   }
   \caption{Achieved contrast ratios for the samples, for a companion detection 5$\sigma$ above both speckle and sky background noise.}
   \label{fig:contrast}
\end{figure}

\begin{table*}
\caption{The wide M-dwarf binary sample.}
\label{tab:targets_hr}
\centering

\begin{scriptsize}
\begin{tabular}{lllllllrrlrc}
\hline
{\bf Name} & {\bf $\bf \rm RA_A$} & {\bf $\bf \rm Dec_A$} & {\bf $\bf \rm SpT_A$} & {\bf $\bf \rm RA_B$} & {\bf $\bf \rm Dec_B$} & {\bf $\bf \rm SpT_B$} & {\bf Dist.[pc]} & {\bf Sep.[AU]} & {\bf Obs. Date} & {\bf Min. sep.[AU]} & {\bf Comp.sep.[AU]}\\
\hline\\
\multicolumn{12}{c}{\textsc{High-resolution $H$ Keck LGS-AO}}\\
SLW1558+0231 & 15 58 51.5  & +02 31 14.7 &       M3.0 & 15 58 53.2  & +02 31 36.9 &       M3.5 &         88 &       3003  &   2008/08/08 &        2.4 &         22$\pm$6 \\
SLW1602+0812 & 16 02 01.4  & +08 12 03.0 &       M2.5 & 16 02 01.0  & +08 11 55.6 &       M4.0 &        153 &       1506  &   2008/08/08 &        6.1 &          \\
SLW1643+2100 & 16 43 56.6  & +21 00 37.8 &       M2.0 & 16 43 56.3  & +21 00 31.3 &       M4.0 &        165 &       1282  &   2008/08/08 &        6.6 &          \\
SLW1714+2807 & 17 14 04.8  & +28 07 05.5 &       M3.5 & 17 14 04.2  & +28 07 04.1 &       M5.0 &        124 &        909  &   2008/08/08 &        4.9 &         10$\pm$2 \\
SLW2137+0032 & 21 37 05.4  & +00 32 31.3 &       M1.0 & 21 37 04.6  & +00 32 46.1 &       M3.5 &        185 &       3473  &   2008/08/08 &        7.4 &          42$\pm$11 \\
SLW2149+0110 & 21 49 47.0  & +01 10 35.1 &       M1.5 & 21 49 47.1  & +01 10 22.0 &       M3.0 &        173 &       2279  &   2008/08/08 &        6.9 &          \\
SLW2156+0017 & 21 56 48.3  & +00 17 00.9 &       M3.0 & 21 56 47.1  & +00 16 53.8 &       M5.0 &        143 &       2807  &   2008/08/08 &        5.7 &          27.2$\pm$6.8 \& 11$\pm$3 \\
SLW2319-0105 & 23 19 53.2  & -01 05 25.4 &       M2.0 & 23 19 52.3  & -01 05 06.5 &       M3.5 &        173 &       4056 
 &   2008/08/08 &        6.9 &          \\
\\
\multicolumn{12}{c}{\textsc{Lower-resolution $K_s$ Keck LGS-AO}}\\
SLW0846+5407 & 08 46 57.7  & +54 07 55.1 &       M3.0 & 08 46 58.7  & +54 07 46.8 &       M4.0 &        132 &       1579  &   2009/04/13 &        7.2 &          \\
SLW0934+3425 & 09 34 52.7  & +34 25 08.7 &       M3.0 & 09 34 53.3  & +34 25 14.4 &       M3.5 &        187 &       1677  &   2009/04/13 &       10.3 &          \\
SLW1022+5835 & 10 22 45.8  & +58 35 58.0 &       M3.5 & 10 22 45.1  & +58 35 50.8 &       M4.0 &         77 &        700  &   2009/04/13 &        4.2 &          \\
SLW1053+0728 & 10 53 09.0  & +07 28 35.2 &       M2.5 & 10 53 08.6  & +07 28 42.8 &       M2.5 &        101 &        923  &   2009/04/13 &        5.6 &          \\
SLW1325+2253 & 13 25 33.3  & +22 53 40.0 &       M4.0 & 13 25 32.4  & +22 53 57.9 &       M4.0 &         67 &       1477  &   2008/05/21 &        3.7 &            \\
SLW1357+1952 & 13 57 14.4  & +19 52 43.1 &       M4.0 & 13 57 13.1  & +19 52 22.5 &       M4.5 &         91 &       2555  &   2008/05/21 &        5.0 &            \\
SLW1406+1546 & 14 06 56.1  & +15 46 32.6 &       M2.5 & 14 06 54.8  & +15 46 05.8 &       M3.5 &        126 &       4186  &   2008/05/21 &        6.9 &           17$\pm$4  \\
SLW1526+2718 & 15 26 48.3  & +27 18 29.8 &       M4.5 & 15 26 49.3  & +27 18 32.8 &       M4.5 &         58 &        798  &   2008/08/08 &        3.2 &            \\
SLW1548+0443 & 15 48 17.7  & +04 43 33.9 &       M3.0 & 15 48 18.3  & +04 43 39.4 &       M3.5 &        137 &       1500  &   2008/08/08 &        7.6 &          $\approx$400 AU (unconf.) \\
SLW1632+1541 & 16 32 13.3  & +15 41 39.0 &       M3.5 & 16 32 12.5  & +15 42 03.8 &       M3.5 &        112 &       3055  &   2008/08/08 &        6.2 &            \\
SLW1639+4226 & 16 39 50.6  & +42 26 12.1 &       M3.0 & 16 39 53.3  & +42 26 15.4 &       M4.5 &        148 &       4426  &   2008/08/08 &        8.1 &          120$\pm$30 \\
SLW2127-0040 & 21 27 49.5  & -00 40 25.5 &       M2.5 & 21 27 51.3  & -00 40 09.7 &       M3.5 &        203 &       6342  &   2008/08/08 &       11.2 &            \\
\\
\multicolumn{12}{c}{\textsc{$K_p$ Palomar 200'' NGS-AO}}\\
SLW1049+6522 & 10 49 08.8  & +65 22 25.2 &       M4.0 & 10 49 09.0  & +65 21 57.4 &       M3.5 &        102 &       2858 &   2009/07/08 &       11.3 &          \\
SLW1323+4429 & 13 23 34.1  & +44 29 55.1 &       M1.5 & 13 23 31.9  & +44 29 49.2 &       M2.5 &        154 &       3827 &   2009/07/08 &       17.0 &           \\
SLW1341-0222 & 13 41 02.5  & -02 22 00.3 &       M3.5 & 13 41 02.8  & -02 21 48.2 &       M4.5 &         52 &        662  &   2009/07/08 &        5.7 &            \\
SLW1345+2857 & 13 45 26.0  & +28 57 23.7 &       M2.5 & 13 45 27.5  & +28 57 24.0 &       M3.0 &        106 &       2159  &   2009/07/08 &       11.7 &          12$\pm$3 \\
SLW1424+1856 & 14 24 15.0  & +18 56 31.7 &       M2.0 & 14 24 14.1  & +18 56 25.6 &       M2.5 &        173 &       2585  &   2009/07/08 &       19.0 &          74$\pm$18  \\
SLW1427+0313 & 14 27 22.6  & +03 13 49.4 &       M3.5 & 14 27 23.5  & +03 13 39.1 &       M3.5 &         85 &       1452  &   2009/07/08 &        9.3 &            \\
SLW1439+5154 & 14 39 32.5  & +51 54 22.1 &       M2.0 & 14 39 33.8  & +51 54 19.2 &       M3.0 &        122 &       1590  &   2009/07/08 &       13.4 &          \\
SLW1500+2306 & 15 00 47.5  & +23 06 26.9 &       M2.0 & 15 00 46.8  & +23 06 51.3 &       M3.0 &        140 &       3684  &   2009/07/08 &       15.5 &            \\
SLW1512+2952 & 15 12 07.1  & +29 52 05.8 &       M3.5 & 15 12 05.9  & +29 52 29.5 &       M3.5 &         60 &       1710  &   2009/07/08 &        6.6 &            \\
SLW1521+5312 & 15 21 26.4  & +53 12 12.4 &       M2.5 & 15 21 27.2  & +53 12 04.4 &       M2.5 &         88 &        967  &   2009/07/08 &        9.7 &            \\
SLW1522+5408 & 15 22 21.2  & +54 08 27.9 &       M1.5 & 15 22 21.1  & +54 08 00.0 &       M3.0 &        120 &       3343  &   2009/07/08 &       13.2 &            \\
SLW1530+3935 & 15 30 21.9  & +39 35 22.1 &       M2.0 & 15 30 22.8  & +39 34 52.7 &       M4.0 &        149 &       4628  &   2009/07/08 &       16.4 &          26.2$\pm$6.6 \\
SLW1624+3249 & 16 24 26.7  & +32 49 57.0 &       M1.0 & 16 24 26.0  & +32 50 00.1 &       M3.0 &        133 &       1373  &   2009/07/08 &       14.7 &           64.2$\pm$16.1 \\
SLW1627+4336 & 16 27 23.6  & +43 36 19.7 &       M1.0 & 16 27 25.1  & +43 36 23.2 &       M3.5 &        160 &       2582  &   2009/07/08 &       17.6 &            \\
SLW1352+4427 & 13 52 18.1  & +44 27 09.7 &       M3.0 & 13 52 18.2  & +44 27 00.7 &       M3.0 &        103 &        929  &   2009/07/08 &       11.3 &            \\
SLW1546-0203 & 15 46 41.5  & -02 03 06.4 &       M4.0 & 15 46 42.3  & -02 02 54.3 &       M5.5 &         55 &        945  &   2009/07/08 &        6.0 &         \\
\end{tabular}
\end{scriptsize}
\tablecomments{Positions are from the SDSS DR7 \citep{Abazajian2009}; spectral types and distances are derived by fitting SDSS and 2MASS photometry using the spectral energy distributions given in \citet{Kraus2007a}. Spectral types are accurate to approximately 1 subclass (extra precision is given to compare populations more easily). Photometric distances are accurate to approximately 0.3 magnitudes, or 15\%. Min. sep. is the minimum detectable separation for an equal-magnitude close companion to either of the star in the target binary. SLW1548+0443A has a companion at a wide radius that requires CPM confirmation (Section \ref{1548_comp}).}
\label{tab:targets}
\end{table*}

\section{Observations and Data Reduction}
\label{obs}

We observed the targets with the Keck II and Palomar AO systems over 4 nights in 2008 and 2009. Keck LGS-AO observations, on 21 May 2008, 8 August 2008, and 13 April 2009, were undertaken in poor conditions with variable cloud cover on all nights. This limited us to the brightest targets of our sample, but did not significantly affect the contrast-performance-limited minimum detectable companion masses. Palomar NGS-AO observations were performed on 8 July 2009 in exceptional seeing, allowing us to use targets as faint as $r$=16.5 as natural guide stars.

We obtained $K_s$-band observations at a depth suitable to detect low-mass brown dwarfs around each target. Stars with binary components detected by eye during observations were also observed in $J$ and $H$ for photometric characterization if conditions and adaptive optics performance allowed. 

The field mid-M-dwarf median binary separation is \mbox{$<$10 AU} \citep{Law2008, Bergfors2010}. At our targets' typical distances of 100 pc, resolving this separation requires the 40 milliarcsecond angular resolution provided by Keck's NIRC2 camera operating in $H$-band with AO. During the majority of the Keck observations conditions were poor, only allowing moderate-Strehl $K_s$ observations, and limiting the minimum detectable separation to $\sim$8 AU. A few hours of the 8 August 2008 night were, however, good enough to push to diffraction-limited resolution in $H$; this subset is large enough to allow us to search for any major systematic effects stemming from a lack of sensitivity to companions between 4 and 8 AU.

The target lists, object properties, image resolutions, and observation epochs are summarized in Table \ref{tab:targets}. Conditions were too poor during the observations of SLW1530+3935B to provide useful limits on its close companion population. However, the primary was well imaged and resolved to be a close binary, and we include this high-order-multiple system in our sample. Most of the data for SLW1624+3249 was lost to an unfortunate disk problem. Data for that target presented here is on the basis of the few frames recovered, although much higher quality images were seen during observations.

All observations were dark subtracted, flat-fielded, sky subtracted and aligned using a custom Python pipeline. The Palomar NGS-AO observations had very variable Strehl ratios due to the extreme faintness of our targets for natural guide star observations. To improve the image quality we took a few tens of 1.4 second exposures for each target. Using the AO frame selection pipeline described in \citet{Law2009}, we aligned each frame using the position of the brightest speckle in the point spread function (PSF), and selected the frames with the highest Strehl ratios to form a final co-added image with much improved resolution.

\subsection{Companion detection}
Binary companions were searched for both automatically and by eye. In almost all cases the two components of the wide binaries were imaged simultaneously, giving excellent PSF references (within the limits of the isoplanatic patch). In other cases, the variable PSFs on these poor nights prevented exact matches with other targets in the survey, although comparison with other stars often did allow rejection of persistent speckles. Three of the detected companions are at a contrast level where speckle noise is significant, and all those targets were imaged simultaneously with the other star in the wide binary. No candidate companions at those flux levels were seen in observations without direct PSF matches.

The maximum contrast ratios for companion detection are shown in Figure \ref{fig:contrast}. The photon and speckle background noise at a particular radius from the primary was determined by: 1) measuring the total flux in each of a set of circular apertures at that radius; 2) calculating the RMS variation in total flux; and 3) requiring a 5$\sigma$ companion detection above the calculated noise. 

\subsection{Companion Photometry}
Aperture photometry was used to determine flux ratios between the components of companions with separations $>$0.3\arcsec. Closer companion flux ratios were derived by an iterative procedure (following \citealt{Kraus2007b}): 1) an estimate analytical point spread function is fit to each component of the binary; 2) the estimated PSF is used to subtract the flux from the secondary; 3) and a new PSF estimate is generated on the basis of the subtracted image. These steps are repeated until convergence, usually in 4-5 iterations. The final estimated PSF is fit to both components simultaneously to derive the flux ratio. Uncertainty estimates noted for the flux ratios include systematic uncertainties from this fitting process.

\section{Results}
\label{results}

\begin{figure*}
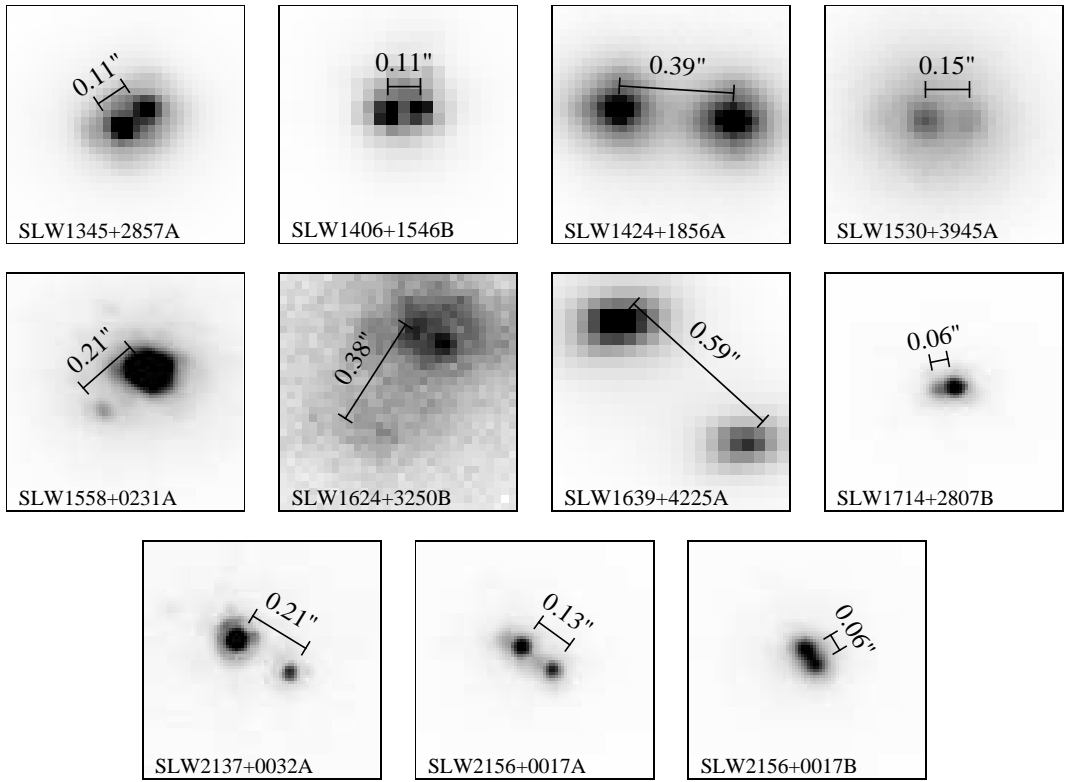

  \centering
	\subfigure{\resizebox{1.25in}{!}{\fbox{\includegraphics{bin_im_1.eps}}}}\hspace{0.15in}
	\subfigure{\resizebox{1.25in}{!}{\fbox{\includegraphics{bin_im_2.eps}}}}\hspace{0.15in}
	\subfigure{\resizebox{1.25in}{!}{\fbox{\includegraphics{bin_im_3.eps}}}}\hspace{0.15in}
	\subfigure{\resizebox{1.25in}{!}{\fbox{\includegraphics{bin_im_4.eps}}}}\hspace{0.15in}

	\subfigure{\resizebox{1.25in}{!}{\fbox{\includegraphics{bin_im_5.eps}}}}\hspace{0.15in}
	\subfigure{\resizebox{1.25in}{!}{\fbox{\includegraphics{bin_im_6.eps}}}}\hspace{0.15in}
	\subfigure{\resizebox{1.25in}{!}{\fbox{\includegraphics{bin_im_7.eps}}}}\hspace{0.15in}
	\subfigure{\resizebox{1.25in}{!}{\fbox{\includegraphics{bin_im_8.eps}}}}\hspace{0.15in}

	\subfigure{\resizebox{1.25in}{!}{\fbox{\includegraphics{bin_im_9.eps}}}}\hspace{0.15in}
	\subfigure{\resizebox{1.25in}{!}{\fbox{\includegraphics{bin_im_10.eps}}}}\hspace{0.15in}
	\subfigure{\resizebox{1.25in}{!}{\fbox{\includegraphics{bin_im_11.eps}}}}\hspace{0.15in}

   \caption{The newly discovered high-order multiple systems. Only the close-binary components are shown. All images are orientated with North up and East to the left and each image is 0.8 arcseconds square. Images are in $H$ with the exception of SLW1530+3945 and SLW1406+1546, both displayed in $K$.}
   \label{FIG:binary_images}
\end{figure*}

Ten of the 36 observed systems were found to contain close companions; one system is a quadruple consisting of two approximately equal-brightness close binaries. The properties of the systems are summarized in Table \ref{tab:discovered_comps} and images of the systems can be found in Figure \ref{FIG:binary_images}.

\begin{table*}
\centering
\caption{The discovered companions}

\label{tab:discovered_comps}

\begin{footnotesize}
\begin{tabular}{llllllllllc}
\hline
{\bf Name}     & {\bf Sep. [\arcsec]} & {\bf P.A. [deg]} & {\bf Dist. [pc]} & {\bf Sep. [AU]} & \bf{$\bf{\Delta K}$} & \bf{$\bf{\Delta H}$}    & \bf{$\rm \bf{(m_{\it K})_{A/B}}$} &{\bf $\rm \bf SpT_{Aa/Ba}$}  & {\bf $\rm \bf SpT_{Ab/Bb}$} & {\bf Wide?}\\
\hline
SLW1345+2857Aa/b  &  0.11$\pm$0.01   & 51$\pm$8      &  110$\pm$30        &  12$\pm$3     & 0.3 $\pm$ 0.1 & 0.31 $\pm$ 0.05 & 10.5  & M2.0                   &     M2.5        &          \\
SLW1406+1546Ba/b  &  0.11$\pm$0.01   &  79$\pm$8   &  150$\pm$40        &  17$\pm$4     & 0.1 $\pm$ 0.2 & \nodata &12.1  & M3.5                   &     M4.0        &      *      \\
SLW1424+1856Aa/b  &  0.39$\pm$0.01   &  276$\pm$2   &  190$\pm$50        &  74$\pm$18    & 0.10 $\pm$ 0.05 & -0.05 $\pm$ 0.05 & 11.5 & M1.5                   &     M2.0        &               \\
SLW1530+3945Aa/b  &  0.16$\pm$0.01   &  94$\pm$5    &  160$\pm$40        &  26$\pm$7     & 1.0 $\pm$ 0.3 & \nodata & 11.5  & M1.5                   &     M3.5        &      *     \\
SLW1558+0231Aa/b  &  0.21$\pm$0.01   & 229$\pm$4     &  100$\pm$30         &  22$\pm$6     & 3.5 $\pm$ 0.3 & 3.5 $\pm$ 0.3 & 11.1  & M2.5                   &     M7.0        &    *     \\
SLW1624+3249Ba/b  &  0.38$\pm$0.05   &  236$\pm$11   &  170$\pm$40        &  64$\pm$16    & 0.7 $\pm$ 0.2 & \nodata & 12.0  & M2.5                   &     M4.0        &      \\
SLW1639+4226Aa/b  &  0.59$\pm$0.01   & 133.8$\pm$1.4     &  200$\pm$50        &  120$\pm$30     & 0.63 $\pm$ 0.05 & 0.69 $\pm$ 0.05 & 12.4  & M3.0                   &     M3.5        &  *         \\
SLW1714+2807Ba/b  &  0.06$\pm$0.01   & 259$\pm$14     &  160$\pm$40        &  9.7$\pm$2.4      & 0.4 $\pm$ 0.2 & 1.46 $\pm$ 0.05 & 13.5  & M4.5                   &     M5.5        &  *         \\
SLW2137+0032Aa/b  &  0.21$\pm$0.01   & 122$\pm$14     &  200$\pm$50        &  42$\pm$11    & 1.60 $\pm$ 0.05 & 1.69 $\pm$ 0.05 & 11.9  & M1.0                   &     M5.0        &            \\
SLW2156+0017Aa/b  &  0.13$\pm$0.01   & 126$\pm$6     &  210$\pm$50        &  27$\pm$7     & 0.4 $\pm$ 0.1 & 0.55 $\pm$ 0.05 & 12.4  & M2.5                   &     M3.5        &   *        \\
SLW2156+0017Ba/b  &  0.06$\pm$0.01   & 149$\pm$13      &  190$\pm$50        &  11$\pm$3     & 0.3 $\pm$ 0.1 & 0.27 $\pm$ 0.05 & 13.7 & M4.5                   &     M5.0        &   *        \\
\end{tabular}
\end{footnotesize}
\tablecomments{$\rm (m_{\it K})_{A/B}$ is the (unresolved) 2MASS K-band magnitude of the close binary system. $\rm SpT_{Aa/Ba}$ refers to the estimated spectral type of the primary component of each close-binary system; $\rm SpT_{Ab/Bb}$ is the estimated spectral type of the newly discovered companion. Contrast ratios are given in $H$ and $K$, where available, and are relative to the nearest member of the target wide binary. The spectral types of the close-companion's primary, the close companion, and the other component of the wide binary are given with an accuracy of approximately one spectral subtype. Systems are noted as ``wide'' if they cross the \citet{Reid2001} empirical mass/separation cutoff. The spectral types and distances have been refit using the resolved photometry, and hence differ slightly from Table \ref{tab:targets}.}
\end{table*}

\subsection{Probability of physical association}
\label{prob_phys_assoc}

Our imaging dataset contains three possible companions at separations $>$3\arcsec, found in a total survey area of 57600 $\rm arcsec^2$. All have SDSS and 2MASS colors inconsistent with being low mass stars at distances similar to our targets, and all are sufficiently bright to have been picked up during any of our observation runs. The sky density of possible companions in our fields is thus a maximum of $\approx$9$\times$$10^{-5}$ stars/$\rm arcsec^{2}$ (at 99\% confidence). Our total companion search area at $<$3\arcsec radius is 2007 $\rm arcsec^2$; we thus expect 0.035 false associations in our dataset, or equivalently a 3\% probability of a single false positive association. This limit is correct for companions with $\Delta M$$\lsim$4 (those detectable in all our observations). Our faintest candidate, SLW1548+0443Ab, has a somewhat lower formal association probability because it would not have been detected in our Palomar imaging, and we leave its confirmation for CPM followup (Section \ref{1548_comp}).

\subsection{Companion Statistics}
\label{binary_fracs}
Eleven of the 71 observed stars are actually close binaries, leading to a raw binary fraction of ${16}^{+5}_{-3}$\% (with a binomial uncertainty estimation). However, our common-proper-motion wide binary selection imposes several biases on the binarity fraction of our target sample: 1) the selection of photometrically-clean, non-extended SDSS objects removes some binaries from the sample; 2) binaries are brighter than single stars at the same colors in unresolved photometry, and so our magnitude-limited sample selects binaries from a larger space volume than the single stars \citep{Burgasser2003}; 3) those binaries found at larger distances are also likely to have decreased proper motion, and so may be removed from our target sample \citep{Law2008}; 4) the widely-separated components of high-order multiple systems containing close pairs can appear to be at different photometric distances in unresolved photometry, removing them from the sample, and 5) the most distant (and therefore the physically widest) binaries will be de-selected by our 7\arcsec minimum-separation cut.

\subsubsection{Biases affecting proper-motion-selected close binaries in the SDSS}
\label{sec:sdss_bins}
We selected our sample from a subset of SDSS sources which are photometrically clean (no error flags), unsaturated, not flagged as galaxies, and also have a clean proper motion determination. Close, barely-resolved binaries in the SDSS may appear as extended sources and so be detected as galaxies, and this could bias our selection away from systems with such components. Similarly, proper motions are only reliable if the nearest neighbouring object in the SDSS is $>$7\arcsec away \citep{Munn2004, Munn2008}; we thus remove objects with nearby resolved companions, and this may similarly bias us against systems with close companions. However, these biases only operate on systems which can be resolved as extended by SDSS.

To determine the maximum detectable close companion separation for our binaries we started with the $<$ 2.5\arcsec separation binaries in the Washington Double Star Catalog \citep{Mason2001}. We then applied our photometric and proper motion selection cuts to those binaries which are present and unsaturated in the SDSS. We find that the SDSS star/galaxy separation does an excellent job of avoiding flagging extended binaries as galaxies -- very few of the binaries are removed at this stage. However, many of the binaries are detected as doubles by SDSS, while additionally some sources have another unrelated object within 7\arcsec (although the removal of those targets does not bias our selection toward or away from binaries). For the purposes of this test we only remove the binaries that have another source within 3\arcsec. This separation is chosen to include our test group of binaries at $<$2.5\arcsec separations (with some margin), but allows us to only flag sources affected by their own (de-blended) companions. 

\begin{figure}
  \centering
  \resizebox{1.0\columnwidth}{!}
   {
	\includegraphics{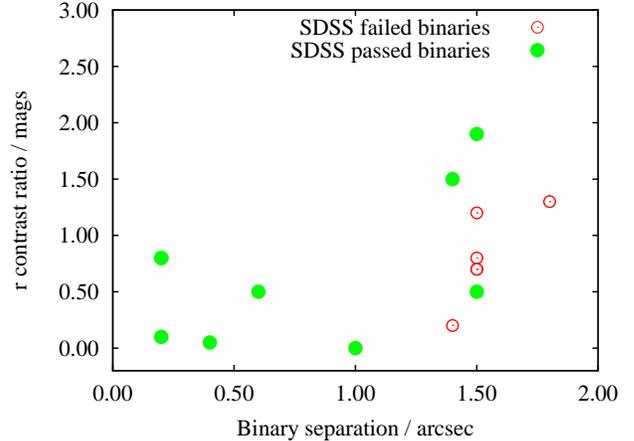}
   }
   \caption{Determining the maximum binary separation which is not removed by SDSS quality cuts. The open circles are binaries which fail either the photometric or proper motion quality cuts due to their close companion; the filled circles are those detected as single stars with good-quality proper motions.}
   \label{fig:sdss_bin_cut}
\end{figure}

The results of this selection are shown in Figure \ref{fig:sdss_bin_cut}. All binaries with $<$1.0\arcsec separation pass the cut (because they are seen as point sources by SDSS), and the higher-contrast binaries at wider separations also pass, because SDSS does not detect the faint companions. Our survey is therefore not significantly biased against binaries up to 1.0\arcsec separation, or $\approx$100AU at our median target distance, and is sensitive to finding new low-mass companions at much larger radii from the target stars (the candidate brown dwarf in Section \ref{1548_comp}, for example).

\subsubsection{Monte-Carlo bias correction}
\label{sec:bias_cor}
To generate correction factors for the other sample-selection biases, we generated an ensemble of simulated wide binary systems in the spectral type, proper motion and distance range of our sample, and then applied our sample selection criteria to produce simulated target lists. This procedure is designed to correct both for the individual biases noted above and for any interaction between the various effects. 

We uniformly distributed 50,000 binary, triple and quadruple systems in a 300pc radius sphere. We then simulated SDSS photometry for each binary component using the spectral energy distributions (SEDs) given in \citet{Kraus2007a}. We included close companions in the generated unresolved photometry, selected from the distributions described below, and included a 2\% photometric uncertainty in the magnitudes. Proper motions for the binaries were taken from a Gaussian distribution with 25$\rm{kms^{-1}}$ width, typical for our thin-disk targets (Section \ref{sample}).

To simulate our sample selection procedure we first obtained a best-fit spectral type and distance for the generated photometry of the binary's components; this step includes the effects of the added unresolved close companions on the inferred spectral types. Because our simulated photometry does not cover the range of SED variations inherent to real stellar populations, we add a 0.3-magnitude Gaussian-distributed uncertainty in the fitted distance modulus. Finally, we apply the distance, proper motion, separation, and magnitude selection criteria discussed in Section \ref{sample}.

We consider a variety of distributions for the spectral types of each component of the multiple systems, by averaging our results between the extremes of a flat distribution between M1 and M4, and simply setting the spectral type to our median spectral type of M2.5. We also average over a range of wide binary separations distributions, between all systems being at 3000 AU, and a flat distribution between 1000 and 9000 AU. Each combination of distributions is also included so, for example, we consider systems with a flat distribution of wide binary component spectral types, but with all close binary components having spectral types equal to their primaries.

Finally, we count the number of selected simulated binary, triple and quadruple systems, and compare them to find the relative size of the ``effective volume'' from which each is taken. This volume directly sets the bias-correction factors for each type of system. The distribution of the close companion spectral types is the only population characteristic that has a $>$10\% effect on the bias-correction factors, with an approximately 2$\times$ change in effective volume between a flat early-M-star distribution and an all-equal distribution. 

We find that triple systems with one unresolved close binary component are suppressed in our sample by between 1.4$\times$ and 3.1$\times$ compared to wide binary systems. This is primarily because a wide binary component that actually consists of two unresolved equal-brightness bodies appears to be closer to us than the other component of the wide binary -- and so is cut out by our aggressive 1$\sigma$ distance match requirement. In contrast, the frequency of selected quadruple systems, not subject to the above effect, is between 0.8$\times$ and 1.2$\times$ the binary fraction, depending on the distribution assumptions. In the high-order-multiple fractions given below, we include the full range of bias-correction factors in our formal uncertainty estimates.

\subsection{Bias-corrected high order multiple fraction}

We discovered 10 high order multiple systems in our 36-target sample, for a raw fraction of ${28}^{+8}_{-6}$\%. 

The bias correction factor for triple and quad systems differs by almost 2$\times$, and the detection of a triple system does not exclude the possibility that it also contains a further unresolved fourth component. However, only one of the 10 high-order-multiple systems is a detected quadruple system. Since quadruple systems have an approximately 2$\times$ greater chance of surviving our sample selection than triple systems, we can interpret this as evidence that quadruple systems are quite rare, at a $\leq$5\% true incidence for our separation range. 

There is an alternative possibility: that a larger fraction of the wide binary systems are in fact quadruple systems, with a small probability that one or both close companions would be at wide enough radius for detection in our AO survey. Some indication of this may be found in the high-resolution subsample, where the only quadruple system in our sample is found with both components at small separations, but a much larger sample is required to fully evaluate this hypothesis. For the purposes of the rest of this paper we take the simplest approach: since 90\% of the detected systems appear to be triples, we assume that the true population of multiple systems is dominated by triple systems. This assumption changes only the absolute value of the high-order-multiple fraction, and does not affect detailed comparisons of multiplicity rates within our sample population.

Adopting the triple bias correction factor from Section \ref{sec:bias_cor} for all 10 multiple systems, and using the binomial distribution probability formalism detailed in \citet{Burgasser2003}, the corrected high-order-multiple fraction is ${45}^{+18}_{-16}$\%.

\subsection{Narrow-angle incompleteness}
We designed the high-resolution Keck $H$ subsample to have sufficient resolution ($<$ 4 AU at 100pc) to probe the 4 AU peak of the field-binary separation distribution of mid-M-dwarfs. Three triple systems and one quadruple system were detected in this 8-system high-resolution subsample, leading to a high-order-multiple fraction of ${70}^{+15}_{-27}$\%, statistically compatible with the binary fraction derived in the lower-resolution portion of the survey. Indeed, only one (20\%) of the companions detected in the Keck high-resolution survey would be undetectable in the lower resolution portion of the surveys. Although these are small number statistics, it appears that insensitivity to companions in the 4-10 AU range does not lead to a large change in the number of detected companions.

\begin{figure}
  \centering
  \resizebox{0.8\columnwidth}{!}
   {
	\includegraphics{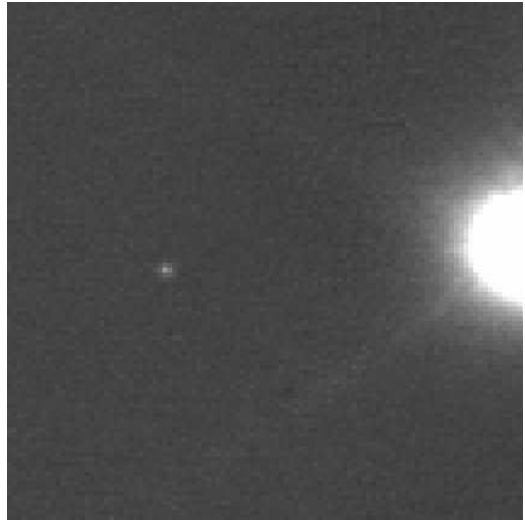}
   }
   \caption{An $H$-band image of the faint possible companion around SLW1548+0443A.}
   \label{FIG:widebd}
\end{figure}

\subsection{Contrast-ratio incompleteness}

\begin{figure}
  \centering
  \resizebox{1.0\columnwidth}{!}
   {
	\includegraphics{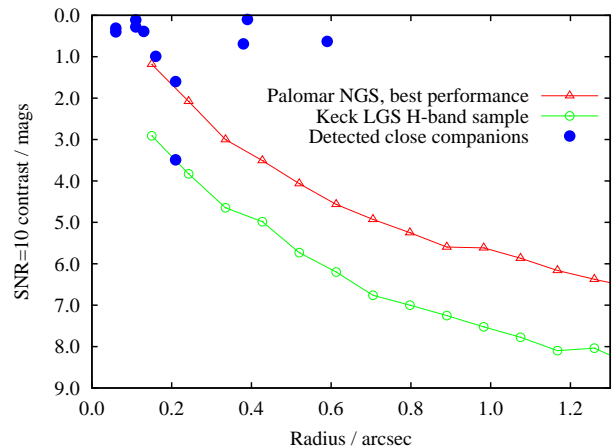}
   }
   \caption{The contrast ratios of the detected companions compared to our detection sensitivities.}
   \label{FIG:contrast_binaries}
\end{figure}

Eight of the eleven close companions are well above our detection sensitivity (Figure \ref{FIG:contrast_binaries}), and they are evenly distributed in the detectable separation range. The remaining three companions are detected just above the sensitivity limit. Although these are clear detections, we cannot exclude the presence of a large population of higher-contrast binaries. 

\subsection{Detected binaries with a high mass ratio}

\subsubsection{A very low mass companion to SLW1558+0231A}

All but one of the confirmed close binaries is approximately equal-mass. The exception, SLW1558+0231Ab, has contrast ratios of $\Delta H=3.4\pm0.25$ and $\Delta K=3.5\pm0.25$ and a projected separation of 22.3$\pm$5.6 AU. Using the \citet{Baraffe1998} models, and our calculated distance modulus for the system, at 5.0 GYr age the secondary is best modelled with a mass of 0.090$\pm$0.007$\rm{M_{\odot}}$ ($1\sigma$ uncertainties). At a 500 MYr assumed age the secondary mass is lowered to 0.080$\pm$0.009$\rm{M_{\odot}}$.

\subsubsection{An unconfirmed wide brown dwarf around SLW1548+0443A}
\label{1548_comp}
A faint object is found just under 3" ($\sim$400AU) away from SLW1548+0443A (Figure \ref{FIG:widebd}). The object is $\sim$7 mags fainter than the target M3 star and is not visible in SDSS because of the high contrast. If the object is associated with the wide M-dwarf binary it has absolute magnitudes $M_J$$\approx$14.9 and $M_K$$\approx$13.7, which is consistent with a mid-mass brown dwarf at the target star distance. Although the probability of association is fairly high, CPM confirmation is required for this companion (Section \ref{prob_phys_assoc}). The system's proper motion is $\approx$0.06\arcsec/$\rm yr^{-1}$, and so it should be possible to confirm the object's association on a short timescale. Since this candidate remains unconfirmed, we do not include it in any subsequent analysis. We will address its association with SLW1548+0443A in more detail in a future paper.

\section{Discussion}
\label{discussion}

\subsection{The Population Statistics of Close Companions in Wide Binary Systems}
The high-order-multiplicity fraction of our sample, ${45}^{+18}_{-16}$\%, is statistically compatible with the 25\% fraction found in the narrower M-dwarf multiple systems in the 8pc sample \citep{Reid2005}. Similarly, the presence of only one detected quadruple system is compatible with the single \citet{Reid2005} quadruple detection.

Our detected companions are almost all at low contrast ratios and relatively small separations on the sky, although two possible very low mass companions were found at larger radii. The distribution of mass ratios is compatible with that found among field close binaries in this mass range \citep{Reid2001, Law2008}.

\begin{figure}
  \centering
  \resizebox{1.0\columnwidth}{!}
   {
	\includegraphics{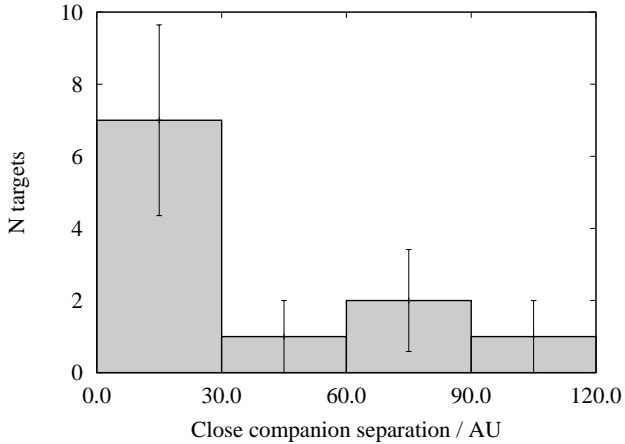}
   }
   \caption{A histogram of the separations of the newly-discovered close companions; 1$\sigma$ Poisson error bars are shown.}
   \label{FIG:comp_sep_hist}
\end{figure}

 The close-binary component separation distribution is strongly peaked at separations $<$30AU (Figure \ref{FIG:comp_sep_hist}); this is three times closer than our $\sim$100AU maximum separation cutoff (see Section \ref{sec:sdss_bins}). At $<$30 AU separations too few systems are detected to allow detailed analysis, but it is worth noting that we do not see see a strong peak towards even closer systems. The survey is sensitive to companions at separations $<$20AU around all our targets, and $<$10AU around two thirds of our targets. A very strong peak towards closer separations, such as that observed in lower-mass M-dwarfs in the field \citep{Law2006, Law2008, Siegler2005}, would be detectable.

\begin{figure}
  \centering
  \resizebox{1.0\columnwidth}{!}
   {
	\includegraphics{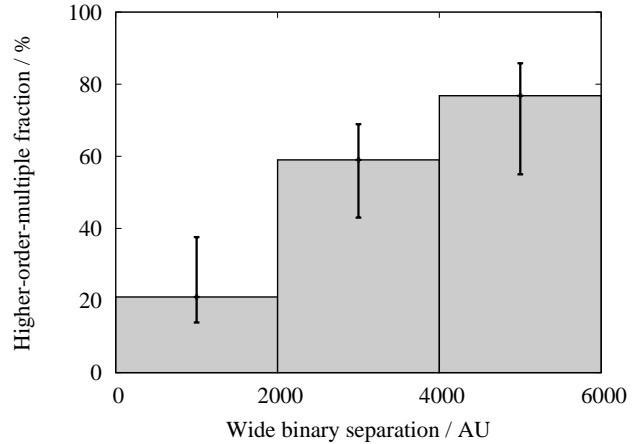}
   }
   \caption{High-order multiplicity as a function of the separation of the wide binary systems. The high-order-multiple fractions, bias corrections, and error bars are calculated as in Section \ref{binary_fracs}, with appropriate Monte-Carlo separation distributions for each of the bins. To compare the different separation ranges without including the range of bias corrections' extra uncertainty in the absolute high-order-multiple fractions, we average over all the bias correction models detailed in Section \ref{binary_fracs}. The final bias corrections at the three separation ranges differ only at the 5\% level.}
   \label{FIG:wide_bin_sep}
\end{figure}

\begin{figure}
  \centering
  \resizebox{1.0\columnwidth}{!}
   {
	\includegraphics{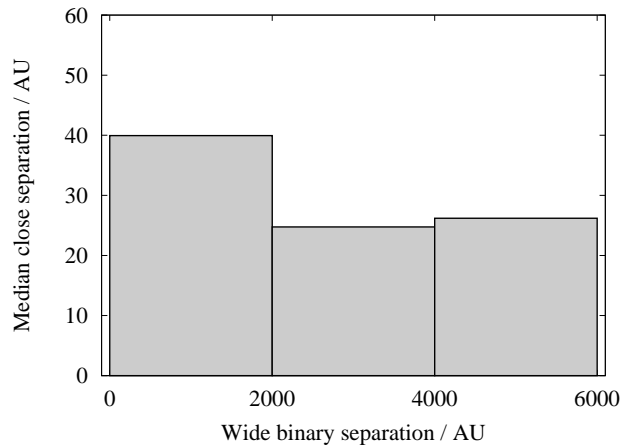}
   }
   \caption{The median close companion separation as a function of wide binary separation. We caution that the leftmost bin contains only two datapoints (10 AU and 64 AU) and should not be interpreted as evidence for a varying median separation.}
   \label{FIG:close_wide_sep_hist}
\end{figure}

\begin{figure}
  \centering
  \resizebox{1.0\columnwidth}{!}
   {
	\includegraphics{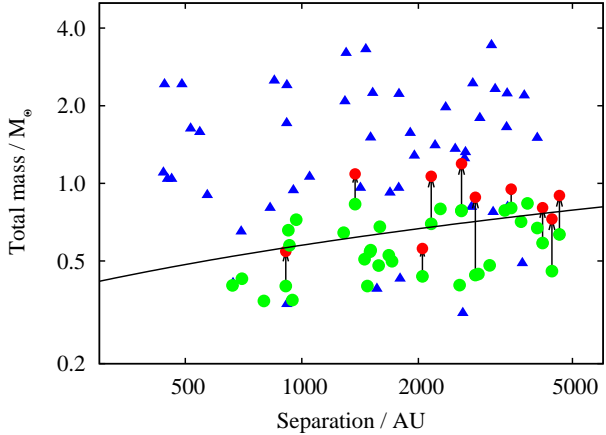}
   }
   \caption{The separations and masses of our wide binary sample (\textit{circles}); our detected high-order multiple systems (\textit{filled circles}) and the young wide binary sample from \citet{Kraus2009} (\textit{triangles}). The effect of binarity on the estimated total system mass is shown for each of our detected higher-order multiple systems, with the lower points based on unresolved photometry and the upper points on AO resolved photometry. The empirical field binary separation cutoff from \citet{Reid2001} is shown as a solid line.}
   \label{FIG:compare_mass}
\end{figure}

\begin{figure}
  \centering
  \resizebox{1.0\columnwidth}{!}
   {
	\includegraphics{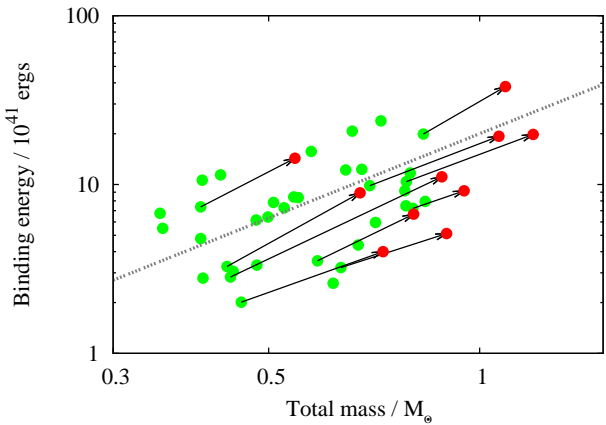}
   }
   \caption{A comparison of the estimated binding energies for our systems before and after the detection of extra components. Initial binding energies from unresolved photometry are shown for all our targets; arrows point to the binding energies after finding additional components. The dashed line is designed to bisect our sample, and has the same binding-energy/mass slope as the Jeans-length-based relations in \citealt{Faherty2010}.}
   \label{FIG:binding_energies}
\end{figure}

\begin{figure}
  \centering
  \resizebox{1.0\columnwidth}{!}
   {
	\includegraphics{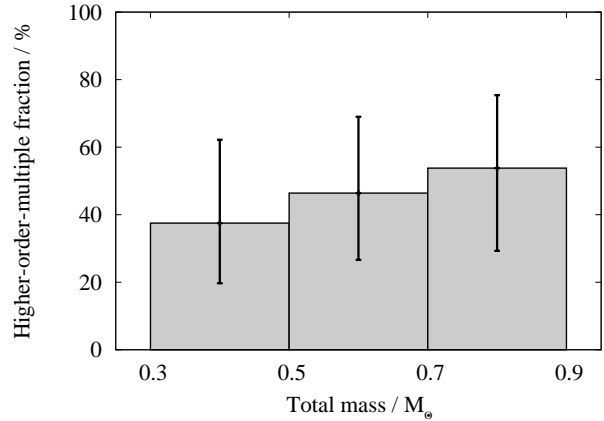}
   }
   \caption{The frequency of high-order multiple systems as a function of total system mass. The system mass of the close binary systems is estimated here from unresolved photometry, for direct comparison with the systems without detected close companions. There is no significant evolution of the high-order-multiple frequency across this mass range.}
   \label{FIG:mass_binary_frac}
\end{figure}

\subsection{High-order Multiplicity}

Next, we assess if the properties of the target wide binaries have an effect on the probability of high-order-multiplicity. In the following discussion, we refer to the newly detected close companions as ``close binaries'', and the systems of which they are part as ``wide binaries''. Binning the targets by the separation of the wide binaries, we find a marginally significant (2$\sigma$) increased high-order-multiple fraction for the wider targets in the survey (Figure \ref{FIG:wide_bin_sep}). For systems with separations up to 2000AU the high-order-multiple fraction is ${21}^{+17}_{-7}$\%, compared to ${77}^{+9}_{-22}$\% for systems with separations $>$4000AU. Indeed, three of the four widest binary targets in the survey are high-order-multiples, while we find only one triple system from eight $<$1000AU systems.

There is no evidence that the median close binary separation changes as a function of the wide binary separation (Figure \ref{FIG:close_wide_sep_hist}). This suggests that the variations in high-order-multiplicity are not an effect of detection efficiency or target selection changing with system distance (as would be expected if uncorrected biases remain in our binarity statistics), but rather are a real change in the multiplicity fraction.

Of the 23 unusually wide systems, six have close-binary components, compared to four close-binary components in 13 systems below the \citet{Reid2001} cutoff (Figure \ref{FIG:compare_mass}). With bias-corrected high-order-multiplicity fractions of ${42.7}^{+21.1}_{-17.1}$\% and ${49.0}^{+22.2}_{-20.7}$\% respectively, the close-binary fractions are statistically indistinguishable. The close-companion mass ratios of the two samples are similarly indistinguishable, while the $22\pm7$ AU median separation of the unusually wide binaries' close-companion population is not significantly different from the $53\pm24$ AU normal-binary median separation. 

Figure \ref{FIG:binding_energies} shows the effect of the resolved companions on the systems' binding energies. Most of our detected high-order-multiple systems are found towards the bottom of the figure, at low binding energies and masses. Although all these systems are above the \citet{Faherty2010} Jeans-length-based binding energy cutoff, it is instructive to consider if there is evidence for the wide binary system properties changing in a similar form to that relation. In Figure \ref{FIG:binding_energies} we also plot a curve with the same power-law slope as the \citet{Faherty2010} relation, with an additional multiplicative factor so that the binding energy cutoff bisects our sample. The full form we adopt is $\rm{E=20M^{5/3}}$, where E is the binding energy in units of $10^{41}$ ergs, and M is the system total mass in solar masses. There are two high-order-multiple systems above this line (a bias-corrected ${20}^{+18}_{-8}$\%), and eight below the line (${62}^{+11}_{-14}$\%). Although we must be cautious to avoid simply picking a relation which fits the data, this marginally signficant difference, and the lack of a similarly significant difference for a simple binding-energy or mass cutoff (Figures \ref{FIG:binding_energies} \& \ref{FIG:mass_binary_frac}), may suggest that Jeans-length considerations are indeed important in the formation of high-order-multiple wide M-dwarf systems. Much larger sample sizes and more detailed statistics are required to fully investigate this.

\subsection{Formation scenarios}

Our results demonstrate that the individual components of unusually wide M-dwarf binary pairs are broadly similar to other low-mass stars in the field, as both populations have similar companion frequencies, separation distributions, and mass ratio distributions (e.g. \citealt{Fischer1992, Siegler2005, Law2008}). 

Less than half of the systems in our observed sample have additional companions at a separation of $\ga$5--10 AU, a comparable rate to typical field M dwarfs.  We are not sensitive to companions at separations of $\la$5--10 AU, so it is possible that the systems have unrecognized additional components inside the detection limits of our survey. However, the absence of a significant additional population at separations of $\sim$5--10 AU in our high-resolution subsample seems to argue against this hypothesis.

The fact that many of our observed systems appear not to contain extra companions (at our detection limits) strongly suggests that an unusually wide separation is not necessarily an indication of (unrecognized) excess mass; some wide binary systems appear to genuinely violate the $\rm{a_{max}-M_{tot}}$ relation seen for most low-mass stars. As this relation was set empirically on the basis of the then-discovered population of M-dwarf binaries, this is not in itself surprising.

However, we do find a 2$\sigma$ increase in the frequency of high-order multiplicity for the widest pairs. This is an intriguing result, as both our survey's smallest-separation and largest-separation binaries would be considered unusually wide compared to the vast majority of field binaries \citep{Reid2001}, and no current models predict that there should be such an evolution in wide binary properties. One possible explanation is that the very widest binaries do in fact need higher total masses to form or to survive, leading to a higher probability of finding extra components in the systems, although it remains to be explained why there would be a rapid change in binary properties at such large separation ranges. Larger sample sizes are required to confirm this observation.

Surveys of star-forming regions show that the frequency of wide ($\ga$500 AU) binary pairs declines smoothly with system mass \citep{Kraus2009}, and we find no evidence that the high order multiplicity of our systems changes with total system mass. It therefore seems plausible that perhaps all protostellar cores undergoing freefall collapse could fragment into multiple systems (as reviewed by \citealt{Bodenheimer2001}) on any length scale, but the probability of this collapse occurring at a given length scale (i.e. the probability of achieving a Jeans-critical overdensity) declines with decreasing characteristic density, and hence with decreasing total mass of the core and the resulting binary system. The two resulting fragments would be otherwise unremarkable, so they would then evolve and further fragment in a manner like low-mass protostars that are not in wide binary pairs. 

This idea would explain both the rarity of wide systems and the
similarity between their components and other low-mass stars. However,
other proposed models could also match the observations, such as the
mutual ejection hypothesis of \citet{Kouwenhoven2010}, which
postulates that wide binary systems could form via simultaneous
ejection from the natal cluster on a similar trajectory. The validity
of such a mechanism should be tested as part of large-scale star
formation simulations (e.g. \citealt{Bate2009, Offner2009}). The increase in high-order-multiplicity for the very widest binaries, if confirmed, could offer a sensitive new test for formation models.

\section{Summary}
\label{conclusions}

We find 11 new close companions in our survey of 36 M-dwarf binaries with separations $>$500AU, giving a bias-corrected high-order-multiple fraction of ${45}^{+18}_{-16}$\% for these wide systems. The close-binary fractions and close-binary mass ratios of our wide binary targets are statistically indistinguishable whether they are above or below the \citet{Reid2001} empirical ``unusually wide'' mass cutoff. We do, however, find marginally significant evidence for an evolution of high-order-multiplicity following the form of the \citet{Faherty2010} Jeans-length binding energy relations. 

The unusually wide binaries' close-companion separations are not significantly different from the normal-binary sample. The close-binary component separation distribution is strongly peaked at separations $<$30 AU, but we do not see see a strong peak towards even closer systems. Almost all the detected companions have similar masses to their primaries, although two very low mass companions, including a candidate brown dwarf, were found at relatively large separations.

We find 2$\sigma$ evidence for an increased high-order-multiple fraction for the widest targets in our survey, with a high-order-multiple fraction of ${21}^{+17}_{-7}$\% for systems with separations up to 2000AU, compared to ${77}^{+9}_{-22}$\% for systems with separations $>$4000AU. Three of the four widest binary targets in the survey are high-order-multiples, while we find only one triple system from eight systems with a wide separation $<$1000AU. It is possible that the widest binaries do in fact need higher total masses to survive, leading to a higher probability of finding extra components in the systems. Larger sample sizes are required to confirm this.

Our results indicate several potentially exciting paths forward in studying unusually wide binary systems. The high frequency of high-order multiples of the widest systems in our sample should be confirmed with larger surveys, and even wider systems should be surveyed to see if the frequency continues to increase. Our sample will also be pushed to much lower total system masses under better LGS observational conditions, and the existing sample should be surveyed with high-dispersion spectroscopy to search for extremely close binary companions.

\acknowledgments
We are particularly grateful to both the Palomar and Keck staff for helping us acquire these targets under difficult conditions. N.M.L. is supported by a Dunlap Fellowship at University of Toronto. S.D., K.G.S., and A.A.W. acknowledge funding support through NSF grant AST-0909463. This research has made use of the SIMBAD database, operated at CDS, Strasbourg, France. 

{\it Facilities:} \facility{Hale, Keck}

\bibliographystyle{apj}
\bibliography{refs}

\label{lastpage}

\end{document}